\documentclass[conference]{IEEEtran}
\IEEEoverridecommandlockouts

\usepackage{cite}
\usepackage{amsmath,amssymb,amsfonts}
\usepackage{algorithmic}
\usepackage{graphicx}
\usepackage{textcomp}
\usepackage{xcolor}

\def\BibTeX{{\rm B\kern-.05em{\sc i\kern-.025em b}\kern-.08em
    T\kern-.1667em\lower.7ex\hbox{E}\kern-.125emX}}

\usepackage[utf8]{inputenc}
\usepackage{booktabs}
\usepackage{tabularx}
\usepackage{makecell}
\usepackage{ragged2e}
\usepackage[draft,bookmarks=false]{hyperref}

\usepackage{multirow}
\usepackage{amsthm}

\begin{document}

\setlength{\textfloatsep}{4pt plus 1.0pt minus 1.0pt}
\setlength{\floatsep}{4pt plus 1.0pt minus 1.0pt}
\setlength{\abovecaptionskip}{2pt}
\setlength{\belowcaptionskip}{2pt}

\setlength{\abovedisplayskip}{2pt}  
\setlength{\belowdisplayskip}{2pt}  
\setlength{\abovedisplayshortskip}{1pt}  
\setlength{\belowdisplayshortskip}{1pt}

\title{Decoding RWA Tokenized U.S. Treasuries: Functional Dissection and Address Role Inference
}


\author{\IEEEauthorblockN{Junliang Luo\IEEEauthorrefmark{1},
Katrin Tinn\IEEEauthorrefmark{2},
Samuel Ferreira Duran\IEEEauthorrefmark{2},
Di Wu\IEEEauthorrefmark{1},
Xue Liu\IEEEauthorrefmark{1}\IEEEauthorrefmark{3}}
\IEEEauthorblockA{
School of Computer Science, McGill University, Montréal, Québec, Canada\IEEEauthorrefmark{1}\\
Desautels Faculty of Management, McGill University, Montréal, Québec, Canada\IEEEauthorrefmark{2}\\
Mohamed bin Zayed University of Artificial Intelligence, Abu Dhabi, UAE\IEEEauthorrefmark{3}\\
\{junliang.luo, samuel.ferreiraduran\}@mail.mcgill.ca, \{katrin.tinn, di.wu5, xueliu\}@mcgill.ca
}
}


\maketitle

\begin{abstract}
U.S. Treasuries have emerged as a prominent subclass of tokenized real-world assets (RWAs), offering cryptographically secured, yield-bearing instruments issued and tradable across multi-(block)chain infrastructures. Why does it matter? It has been argued that tokenizing mainstream financial assets brings value via transparency, accessibility, and financial inclusion. 
While the market has expanded rapidly, empirical analyses of transaction-level behaviors remain limited. 
This paper conducts a quantitative, function-level dissection of U.S. Treasury-backed RWA tokens, including BUIDL, BENJI, and USDY, across multiple chains: mostly Ethereum and Layer-2s.
Decoded contract calls are used to identify core financial primitives such as issuance, redemption, transfer, and bridging, revealing patterns that distinguish institutional participants from trades on the secondary market, possibly by smaller and retail users.
To infer address-level economic roles, we introduce a curvature-aware representation learning model. 
Our method outperforms baseline models in role inference on our collected U.S. Treasury transaction dataset and generalizes to address classification across broader public blockchain transaction datasets.
The decoded transaction-level patterns in tokenized U.S. Treasuries across chains unveil the degree of secondary-market trading. Our inference model enables the distinction between institutional investors and issuers, arbitrage bots, and other traders based on behavioral patterns. 
\end{abstract}


\section{Introduction}
\label{sec:introduction}
Tokenization of real-world assets on blockchain means representing real-world assets on-chain through tokenized representations, potentially enhancing enforceability and tradability \cite{ledger2024rwa}. 
An increasingly prominent form of such tokenization involves yield-bearing financial instruments such as sovereign debt, Treasuries, credit products, and real estate, represented on-chain as cryptographically enforceable claims.
While such arrangements involve complications related to compliance-bound issuance frameworks and asset custody infrastructures \cite{scharfman2021cryptocurrency}, RWA tokens have evolved to encompass a broader class of fixed-income and yield-generating instruments, notably short-duration U.S. Treasuries \cite{riabokin2024role}, tokenized real estate \cite{smith2025real}, and private credit loans \cite{alamsyah2023revealing}, with each implemented through on-chain contracts governed by legal wrappers, custodial attestations, and programmable functions embedded in ERC-20 \cite{bauer2022erc} or ERC-1400 \cite{securitytokenstandard} token standards.
Such tokenized assets are increasingly viewed through the lens of financial inclusion, particularly for enabling access to sovereign debt instruments across geopolitical and socioeconomic boundaries, e.g., allowing individuals in emerging markets or lacking U.S. broker accounts to access U.S. Treasuries via on-chain platforms.
Tokenized representations of short-duration U.S. Treasury bills have emerged as a dominant subclass, expected to reduce issuance or settlement frictions and compliance costs, as well as broaden global (including retail) access to mainstream securities \cite{channing2024tokenization}. 
Such instruments complement DeFi as safe collateral and programmable primitives \cite{alamsyah2024review}, constituting over \$6.1 billion in on-chain assets as of Q2 2025 \cite{rwa2025treasuries}.
Among leading instruments, we focus on BUIDL (BlackRock), BENJI (Franklin Templeton), and USDY (Ondo): the largest by market capitalization and on-chain activity as of Q2 2025 \cite{rwa_treasuries_2025}.
Functionally, these tokens implement yield accruals via rebasing mechanics \cite{ondo2025ousg} or dividend distribution \cite{steakhouse2024buidl}, and embed smart contracts that encode the logic and enforce verifiable transfers.
Despite the market’s growth, it remains unclear whether such instruments meaningfully expand access beyond institutional actors to retail participants. The transparency of on-chain behavior in shaping an inclusive and accountable financial infrastructure remain insufficiently understood.
Prior research on RWA tokens centred only on features of asset issuance and transfer mechanisms \cite{baltais2024economic, chen2024exploring}, institutional governance trade-offs \cite{tanveer2025tokenized}, composability limitations \cite{chen2024exploring}, and participation frictions \cite{swinkels2023empirical, kreppmeier2023real}.
To date, no studies have extracted transaction-level data to construct function-decoded, cross-chain token transfers and conduct address role modeling, the potentially foundational components for analyzing whether RWA tokenization benefits more equitable access financial systems, and broader socioeconomic inclusion in line with Sustainable Development Goals \cite{assembly2015sustainable}.
Concretely, three critical deficiencies: The initial challenge is dataset construction: no public resource has aggregated a cleaned, formatted collection of an RWA U.S. Treasury transaction dataset, with normalized chain-specific metadata, and supplied tractable address (role) labels across multiple blockchains. 
The absence of such a processed, multi-chain dataset precludes reproducible cross-chain analytics and limits empirical validation of claims. 
Secondly, no prior study has undertaken a transaction-level functional audit of RWA U.S. Treasury token contracts: The transactions have not been decomposed into economic primitives such as issuance, redemption, liquidity provision, bridging; nor have the analyzed results that have been traced to insights of research questions including (i) quantify institutional versus retail participation; which group exhibits broader engagement in the market. (ii) characterize how individual chains specialise within the multi-blockchain deployment stack, clarifying why specific chain hosts some distinct financial functionalities.
In addition, the literature lacks a principled weak-supervision model capable of mapping on-chain addresses to behavioral roles (e.g., custodian, treasury, execution bot, trader), thereby limiting downstream applications such as compliance screening \cite{wohlgemuth2019competitive}, AML surveillance \cite{aidoo2025role}, and liquidity-risk attribution \cite{dixon2019blockchain}. Such modeling also strengthens transparency and accountability in blockchain-based financial systems, advancing infrastructure aligned with goals related to inclusive economic growth and reduced inequalities \cite{assembly2015sustainable}.
To our knowledge, this is the first study to systematically structure transaction-level data for tokenized U.S. Treasuries, enabling quantitative, function-level analysis.
Our contributions are as follows:

\noindent\textbf{U.S. Treasury Tokens Transaction Dataset}: We assemble a multi-chain transaction dataset for RWA U.S. Treasury tokens, including BUIDL, BENJI, and USDY, mapping verified contracts, decoding method calls, and collecting community-accumulated address annotations with simplified role labels to enable reproducible, function-level analysis; the dataset will be made available upon publication.

\noindent\textbf{Cross-chain functional decomposition}: We present a contract-decoded, function-level analysis of tokenized Treasuries across multiple chains, revealing how issuance, redemption, transfer, and DeFi-related operations vary by chain and token design. 
We perform statistical segmentation of wallet behaviour using transaction frequency and value distributions, identifying the differences between the institutional and retail actors. We find that institutional wallets dominate issuance and redemption flows, while retail users engage primarily in mid-value transfer activity concentrated on Layer-2 networks.

\noindent\textbf{Address role inference model}: We propose a predictive model based on Poincaré embeddings to capture latent transaction geometry for address role inference (e.g., treasury, bots, traders). Our approach outperforms established baselines on the RWA Treasury dataset and exhibits competitive generalization to downstream tasks such as anomalous address classification on diverse public blockchain transaction network datasets.

\section{Related Work}
Real-world asset tokenization has increasingly focused on fixed-income instruments, with recent studies examining blockchain-based representations of sovereign debt, particularly tokenized U.S. Treasuries, as a representative use case within on-chain capital markets \cite{baltais2024economic}.
A recent multi-sector case study similarly reports that tokenization can improve transaction efficiency and create new value, yet it also introduces governance complexities and shifts risk distribution in on-chain markets \cite{tanveer2025tokenized}.
In addition, empirical surveys of thirty-nine RWA projects have been studied to reveal common on-chain vulnerabilities, such as heavy reliance on stablecoins for settlement and a limited base of active on-chain investors \cite{chen2024exploring}. 
From a behavioural perspective, research efforts have been undertaken to examine address-level data to characterize RWA usage patterns. 
Swinkels studied Ethereum-based real estate tokens (fractional shares of 58 U.S. rental properties) and found highly fragmented ownership, with around 254 unique holders per property on average \cite{swinkels2023empirical}. 
The author observed that larger token holders tended to spread investments across multiple properties, while overall liquidity was low: each property token changed ownership only about once per year (slightly more often if tradable on decentralized exchanges). 
Kreppmeier et al. tracked 173 U.S. real estate security tokens (over 238,000 on-chain transactions) and similarly found broad participation by small investors, though individual wallets were not well-diversified across different tokens \cite{kreppmeier2023real}. 
In addition to property fundamentals, the study showed that crypto-market factors such as transaction costs and investor sentiment significantly influence both initial token offering success and subsequent trading flows. 
These studies demonstrate the value of categorizing on-chain addresses by their roles and behaviours, for instance, distinguishing issuers, custodians, and various types of investors to better interpret RWA economical patterns. 
The emerging of U.S. Treasury RWA tokens, i.e., on-chain shares in Treasury-backed funds, has drawn interest for bringing safe assets on-chain \cite{riabokin2024role}, although academic analysis of their transaction-level semantics such as participant composition remains scant.

\section{U.S. Treasury Tokens Dataset Collection}
\label{sec:data-collection}
We collect raw on-chain transactions for tokenized real-world asset products representing U.S. Treasuries, namely BUIDL (BlackRock USD Institutional Digital Liquidity Fund), BENJI (Franklin OnChain U.S. Government Money Fund), and USDY (Ondo U.S. Dollar Yield) across multiple blockchains using public blockchain explorer APIs.
Our collection pipeline queries EVM-compatible chains (e.g., Ethereum, Arbitrum, Mantle), as well as non-EVM chains (e.g., Aptos, Stellar), and fetches transfers for the three most prominent RWA tokens by market capitalization: BUIDL, BENJI, and USDY based on rankings from market cap \cite{rwa_treasuries_2025}.
We enumerate and resolve all verified smart contract deployments across their supported chains, extracting token transfer activity via the following official block explorers: \textit{Etherscan} (Ethereum), \textit{Polygonscan} (Polygon), \textit{Arbiscan} (Arbitrum), \textit{Basescan} (Base), \textit{Optimistic Etherscan} (Optimism), \textit{Snowtrace} (Avalanche), and \textit{Aptoscan} (Aptos). Transaction logs were retrieved from contract inception through April 2025, spanning 11 chain-token pairs.

\begin{table}[t]
\centering
\caption{Cross-chain summary statistics for tokenized U.S. Treasury. Columns report the number of transactions, unique participating addresses, and the range of observed activity.}
\renewcommand{\arraystretch}{0.85}
\label{tab:rwa-data-summary}
\resizebox{0.48\textwidth}{!}{%
\begin{tabular}{llrrll}
\toprule
\textbf{Token} & \textbf{Chain} & \textbf{Transaction} & \textbf{Address} & \textbf{Start} & \textbf{End} \\
\midrule
\multirow{4}{*}{\textbf{USDY}} 
  & Arbitrum  & 13,296  & 1,281 & 2024-08-12 & 2025-04-28 \\
  & Aptos     & 10,000  & 2,259 & 2024-08-31 & 2025-04-22 \\
  & Ethereum  & 2,921   & 729   & 2023-09-18 & 2025-04-21 \\
  & Mantle    & 274,657 & 7,221 & 2023-12-23 & 2025-04-22 \\
\midrule
\multirow{7}{*}{\textbf{BENJI}} 
  & Polygon   & 996     & 5     & 2023-10-03 & 2025-04-23 \\
  & Arbitrum  & 1,112   & 7     & 2023-11-13 & 2025-04-25 \\
  & Base      & 289     & 4     & 2024-11-20 & 2025-04-23 \\
  & Avalanche & 356     & 5     & 2024-10-11 & 2025-04-23 \\
  & Aptos     & 142     & 3     & 2024-10-01 & 2025-04-23 \\
  & Ethereum  & 212     & 3     & 2024-11-20 & 2025-04-23 \\
  & Stellar   & 2,546,750 & 220,784 & 2024-02-22 & 2024-03-06 \\
\midrule
\multirow{5}{*}{\textbf{BUIDL}} 
  & Ethereum  & 4,639   & 65    & 2024-03-04 & 2025-04-21 \\
  & Polygon   & 204     & 7     & 2024-11-04 & 2025-04-21 \\
  & Arbitrum  & 135     & 6     & 2024-11-04 & 2025-04-21 \\
  & Optimism  & 136     & 4     & 2024-11-04 & 2025-04-21 \\
  & Avalanche & 25      & 6     & 2024-11-04 & 2024-12-06 \\
\bottomrule
\end{tabular}
}
\renewcommand{\arraystretch}{1.0}
\end{table}

BUIDL shows the most activity on Ethereum, aligning with its institutional adoption, though it is collected across multiple chains. The BENJI token exhibits exceptionally high activity on Stellar, with over 2.5 million transactions involving more than 220,000 addresses in just two weeks, indicating high-frequency issuance or custodial activity. 
USDY shows dense activities on Mantle, accounting for more than 274,000 transactions. 
Overall, this collected multi-chain dataset enables us to quantify RWA adoption patterns and behavioural clusters in later sections.
To decode smart contract method calls, we utilize Tenderly's decoding engine \cite{tenderly2025} and the open-source 4byte signature database \cite{4byte_ethereum} to decode the \textit{input} field of each transaction, enabling function-level analysis. The result is a cross-chain, function-decoded dataset of RWA token transfers suitable for downstream quantitative analysis.

\section{Quantitative Analysis of RWA Tokens}

To quantitatively assess whether U.S. Treasury token transactions align more closely with institutional investors or retail users, we analyze the frequency of transactions versus their size.
An institutional usage pattern is characterized by relatively few transactions, each of very large notional value, while a retail-driven pattern involves a high frequency of smaller transactions, typically ranging from a few dollars to several thousand.
We further stratify this analysis by both chains and functions to assess how usage patterns vary across blockchain environments (e.g., Ethereum vs. Arbitrum) and operational roles (e.g., issuance, redemption, transfer).

\begin{figure}[b!]
    \centering
    \includegraphics[width=0.49\textwidth]{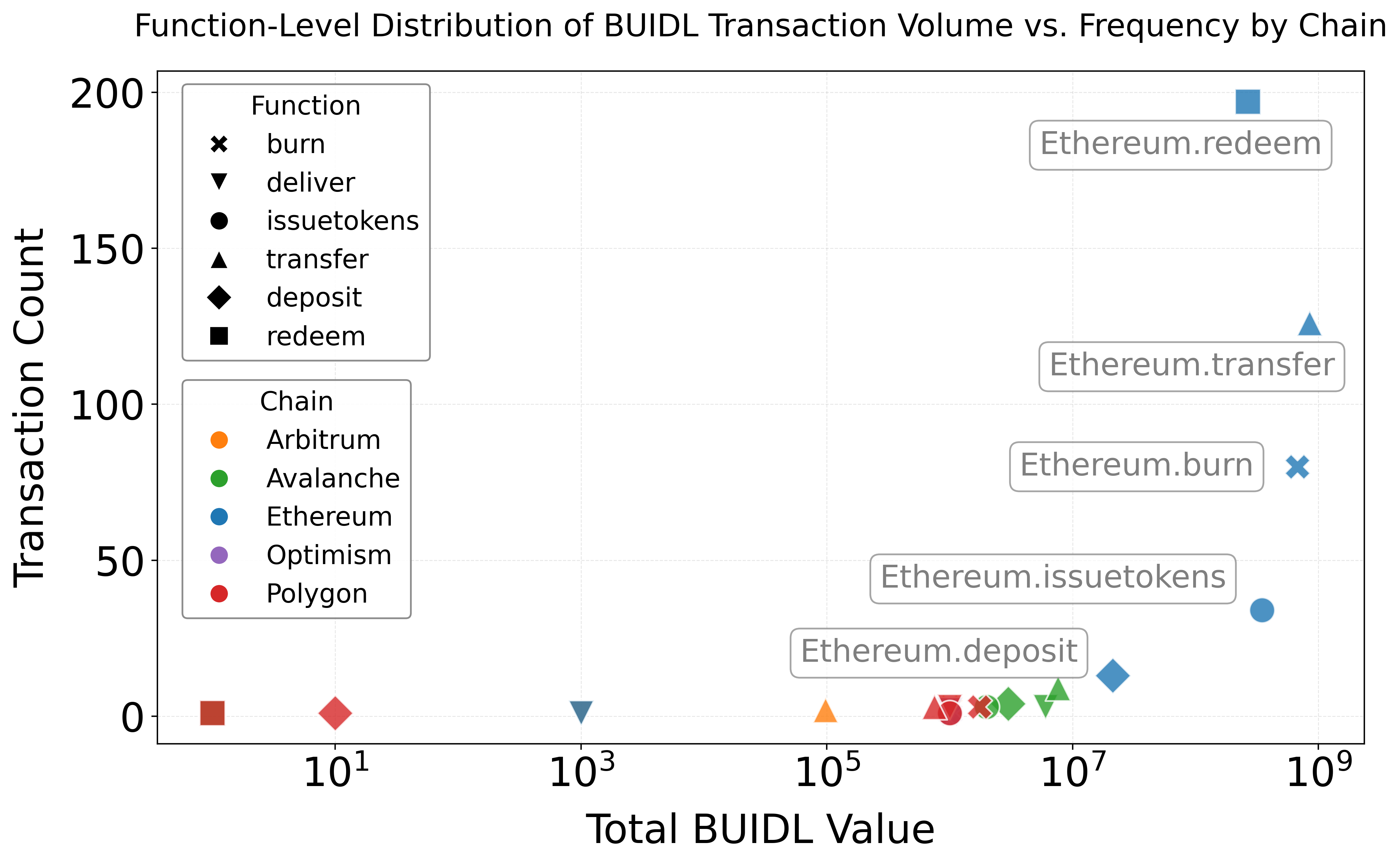}
    \caption{
    Log-scaled scatter plot of BUIDL token transactions by function and chain. Each point represents a \textit{(function, chain)} pair, where the x-axis is the total BUIDL value transacted and the y-axis is the transaction count. Marker shape denotes functional class (e.g., \textit{transfer}, \textit{issuetokens}, \textit{redeem}), and colour indicates a blockchain (e.g., Ethereum, Arbitrum). }
    \label{fig:buidl-function-volume-vs-frequency}
\end{figure}

\begin{table}[h]
\centering
\caption{Functional buckets for decoded contract functions. Substrings are matched against decoded function names. Functions reflect high-level operations in RWA token flows.}
\label{tab:functional-buckets}
\resizebox{0.48\textwidth}{!}{%
\begin{tabular}{lll}
\hline
\textbf{Func buckets} & \textbf{Name Match} & \textbf{Description} \\ \hline
\textit{issuetokens} & issue, mint, bulkissuance & Initial minting and dividend payments \\
\textit{transfer} & \begin{tabular}[c]{@{}l@{}}transfer, bridgedstokens, \\ multisend\end{tabular} & \begin{tabular}[c]{@{}l@{}}Routine transfers or bridging \\ between chains\end{tabular} \\
\textit{redeem} & redeem & Investor withdrawals \\
\textit{burn} & burn & Permanent destruction of tokens \\
\textit{deposit} & deposit & Vault top-ups or liquidity provisioning \\
\textit{deliver} & deliver & Fee or metadata delivery helpers \\ \hline
\end{tabular}
}
\end{table}

\noindent \textbf{\textit{BUIDL}}.
We categorize the contract decoded functions into operational buckets based on string-matching heuristics against function names to interpret the functional semantics. 
Table~\ref{tab:functional-buckets} summarizes these mappings, which allow for cross-chain, token-agnostic aggregation of function activity into economically meaningful categories. 
For instance, the \textit{issuetokens} bucket includes all minting operations ranging from initial token creation to dividend payments (e.g., via batched bulkIssuance calls in BUIDL).
Likewise, \textit{redeem} captures capital outflows by investors, while \textit{burn} reflects permanent supply contraction. 
The \textit{transfer} bucket encompasses routine value transfers, including bridged hops and multi-send operations. 
Figure~\ref{fig:buidl-function-volume-vs-frequency} visualizes the distribution of BUIDL token interactions by function and chain, where each data point corresponds to a unique $(f, c)$ pair: a specific function (bucket) $f$ executed on chain $c$. 
The x-axis (log scale) captures the total value transferred through the function, while the y-axis captures the number of transactions. Marker shapes denote function classes (e.g., \textit{transfer}), and colours distinguish the underlying chains (e.g., Ethereum, Polygon, Arbitrum).
Points located in the upper-right quadrant represent functions with both high frequency and high cumulative value. 
In contrast, points near the upper-left indicate high-frequency, low-value interactions consistent with retail-like activity, and points in the lower-right denote low-frequency but high-value functions, as expected for capital-intensive institutional transfers.
The observed distribution of BUIDL's on-chain activity is heavily skewed toward the right, especially on Ethereum, with large-value function calls such as \textit{redeem}, \textit{transfer}, \textit{burn}, and \textit{issuetokens} dominating the volume. 
This pattern is consistent with usage by accredited or institutional actors, rather than mass-market retail adoption. The dominance of Ethereum reinforces its role as the canonical execution environment for regulated, institutional-grade RWA issuance and settlement.
However, the presence of BUIDL transactions on alternative chains such as Avalanche, Polygon, and Optimism indicates emerging interoperability requirements. These deployments may support secondary custody flows, bridging infrastructure, or protocol-level integrations where gas efficiency or modular composability is prioritized. 
Such cross-chain activity for BUIDL remains limited as of April 2025, based on our collected data, with the vast majority of transaction volume and functional engagement concentrated on Ethereum.
Notably, BUIDL transactions are almost exclusively initiation-side operations (e.g., primary issuance and redemption), with minimal secondary trading activity, suggesting that most interacting addresses represent institutional counterparties rather than retail participants.
%

\noindent \textbf{\textit{BENJI}}.
Franklin Templeton’s BENJI token representing shares of its OnChain U.S. Government Money Fund, was initially deployed on the Stellar blockchain in 2021 and has since used Stellar as the primary ledger for recording share ownership and transactions \cite{stellar2023benji}.
From 2023 to 2024, BENJI was expanded beyond Stellar onto several EVM-compatible chains: Polygon, Arbitrum, Avalanche, and Ethereum.
Each blockchain of the BENJI token’s plays a specialized role. 
Stellar still functions as the primary issuance and settlement network, where the majority of BENJI tokens are minted and held (most of the token supply was on Stellar as of April 2024 \cite{gilbert2024benji}).

\begin{table}[htbp]
\centering
\caption{
Transaction summary for BENJI across supported chains and decoded functions, based on data collected through April 2025. Stellar shows exceptionally high activity due to its role as the fund’s primary settlement ledger, while EVM chains primarily use the \textit{signedDataExecution} function for controlled, authorized operations.
}
\label{tab:benji-signeddata}
\resizebox{0.49\textwidth}{!}{%
\begin{tabular}{l l r r r l l}
\toprule
\textbf{Chain} & \textbf{Function} & \textbf{Transaction} & \textbf{Total Value} & \textbf{First Seen} & \textbf{Last Seen} \\
\midrule
Arbitrum  & signeddataexecution & 622      & 192{,}720{,}281    & 2023-11-13 & 2025-04-25 \\
Avalanche & signeddataexecution & 353      & 34{,}645{,}861     & 2024-10-11 & 2025-04-23 \\
Base      & signeddataexecution & 185      & 31{,}521{,}747     & 2024-11-20 & 2025-04-23 \\
Ethereum  & signeddataexecution & 185      & 30{,}459{,}341     & 2024-11-20 & 2025-04-23 \\
Polygon   & signeddataexecution & 938      & 30{,}640{,}118     & 2023-10-03 & 2025-04-23 \\
Stellar   & -            & 2{,}546{,}750 & -         & 2024-02-22 & 2024-03-06 \\
\bottomrule
\end{tabular}}
\end{table}

The BENJI token’s on-chain activity demonstrates a stark bifurcation between Stellar and EVM-compatible chains. 
On Stellar, over 2.5 million transactions were recorded within 14 days, primarily driven by operational activity such as dividend reinvestments, account initializations, and share registry updates. 
These actions occur at high frequency and low value per transaction, and although their financial values are not directly embedded in the transaction logs, they are encoded in structured parameters that require further decoding.
On Stellar, transactional values are not directly recorded in the standard fields but are embedded within parameters of the function metadata, requiring further decoding for precise financial analysis.
In contrast, all detectable transactions on EVM chains: Ethereum, Polygon, Arbitrum, Avalanche, invoke a single function: \textit{signedDataExecution}. 
This meta-transaction handler encapsulates pre-authorized operations (e.g., mint, transfer, redeem) within a signed payload, enabling BENJI to enforce off-chain compliance and centralized control over all state changes on EVM chains.
%

\noindent \textbf{\textit{USDY}}.
We also categorize USDY smart contract functions into operational buckets by applying heuristic string matching on the decoded function names. 
Table~\ref{tab:usdy-functional-buckets} describes this mapping. 
Given the variety of protocols and contracts interacting with USDY across multiple chains, the mapping produced an aggregation based on the functional semantics. 
For example, swap captures all decentralized exchange operations regardless of type, while lending encompasses credit-related actions such as borrow, repay, or collateral management. 

\begin{figure}[htbp!]
    \centering
    \includegraphics[width=0.49\textwidth]{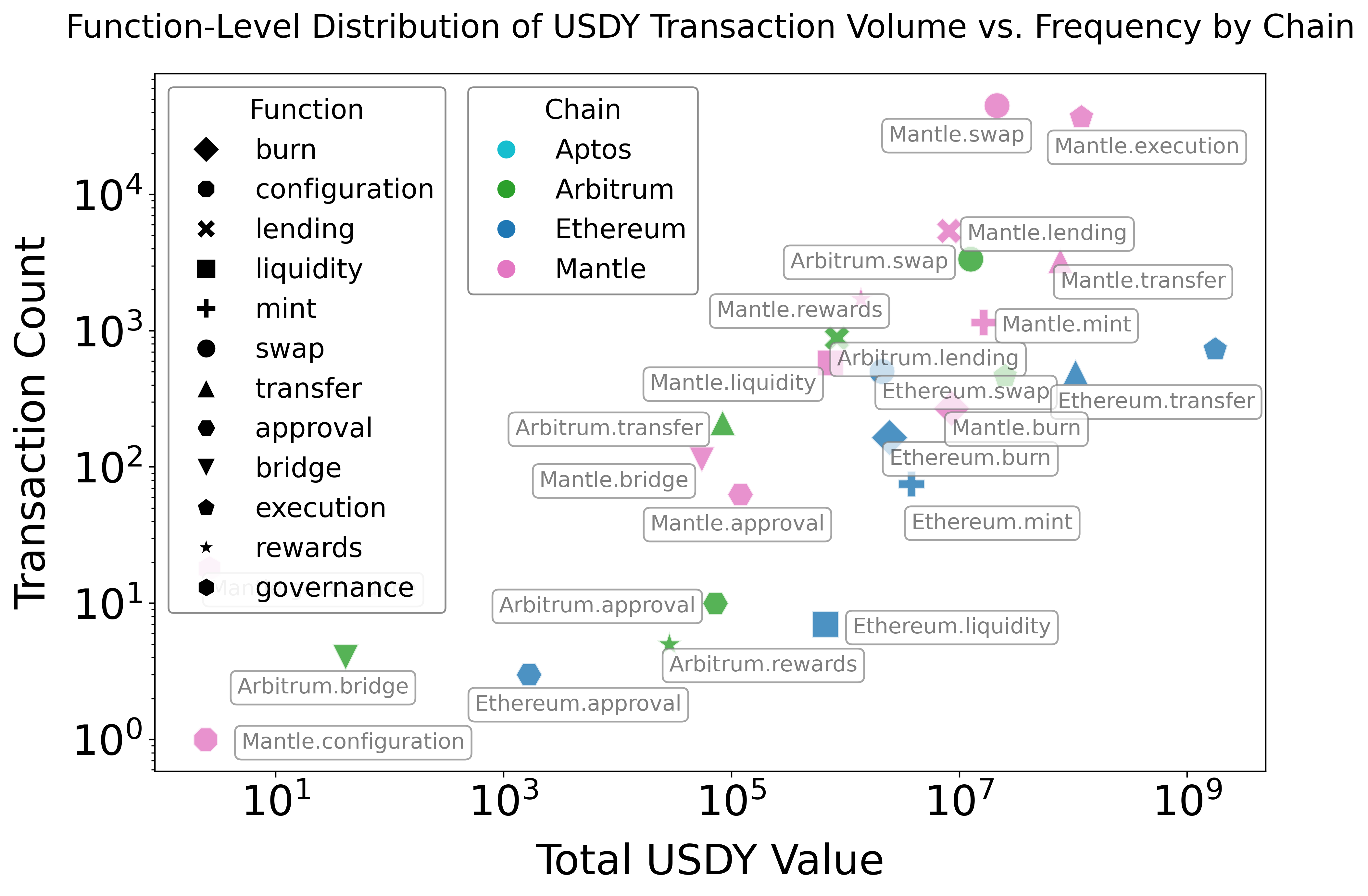}
    \caption{Function-level distribution of USDY transaction volume vs. frequency across chains. 
    Each point represents a specific $(f,c)$ pair (function $f$ on chain $c$), with marker shape encoding functional category (e.g., swap, lending, execution) and colour denoting blockchain. 
    Both axes use logarithmic scales. This figure reveals how USDY’s cross-chain activity exhibits distinct clusters of functional usage, highlighting protocol specialization (e.g., high-frequency swaps on Mantle vs. large-value mint and burn operations on Ethereum).}
    \label{fig:usdy-function-chain}
\end{figure}

\begin{table}[h]
\centering
\caption{Functional buckets for USDY contract activity. Buckets are derived by substring pattern matching in decoded function names. These categories reflect common DeFi roles in cross-chain real-world asset tokens.}
\label{tab:usdy-functional-buckets}
\resizebox{0.49\textwidth}{!}{%
\begin{tabular}{lll}
\hline
\textbf{Func Bucket} & \textbf{Name Match (substring)} & \textbf{Description} \\ \hline
\textit{swap} & swap, unoswap, swaptoken & Token swaps via routers or aggregators \\
\textit{liquidity} & add\_liquidity, removeliquidity & Liquidity pool provisioning or withdrawal \\
\textit{lending} & lend, borrow, repay, loan, collateral & Lending, borrowing, collateral adjustment \\
\textit{transfer} & transfer, transfertoken, safetransfer & Standard token transfers between addresses \\
\textit{bridge} & \begin{tabular}[c]{@{}l@{}}bridge, startbridge, \\ swapandstartbridge\end{tabular} & Cross-chain bridging of tokens \\
\textit{mint} & mint & Token issuance, typically from off-chain trigger \\
\textit{burn} & burn & Token removal or redemption \\
\textit{rewards} & claim, harvest, reward, collect & Claiming rewards or accrued yield \\
\textit{governance} & vote, governance & Governance interactions (e.g., DAO voting) \\
\textit{approval} & approve, permit & Token approvals or permission signatures \\
\textit{configuration} & register, set\_, init, config & Administrative or protocol configuration actions \\
\textit{execution} & \begin{tabular}[c]{@{}l@{}}executemeta, execute, \\ exectransaction, \\ delegatecall, call, multicall\end{tabular} & \begin{tabular}[c]{@{}l@{}}General-purpose call wrappers and \\ execution shells\end{tabular} \\
\textit{unknown} & (none matched) & Uncategorized or obscure logic \\ \hline
\end{tabular}
}
\end{table}

Figure~\ref{fig:usdy-function-chain} visualizes the cross-chain function-level distribution of USDY transactions by plotting total token value against transaction count for each \textit{(function, chain)} pair. 
The on-chain data for USDY reveals a mix of institutional-sized wallets and numerous retail holders.
The function-level transaction data for USDY reveals a pattern indicative of both institutional and retail engagement across multiple blockchains. 
Despite USDY's positioning as a regulated, yield-bearing stablecoin offered under Regulation S \cite{regulationS_} to non-U.S. investors, on-chain evidence suggests that the asset has achieved a relatively broad distribution: over 10,000 unique addresses hold USDY across supported chains, suggesting significant uptake by smaller-scale retail participants.
However, transaction size and function type diverge markedly between user classes. Large-value, low-frequency operations such as \textit{mint}, \textit{burn}, and high-value \textit{execution} calls are primarily concentrated on Ethereum, consistent with institutional-scale issuance and redemption flows. 
These interactions exhibit transaction sizes in the hundreds of thousands to millions of USDY, aligning with primary market activity and custody-level fund management.
By contrast, Layer-2 networks such as Arbitrum and Mantle exhibit distinct transactional profiles characterized by higher frequency, lower median value, and increased heterogeneity in function types (e.g., \textit{swap}, \textit{transfer}, and permissioned \textit{execution}). These patterns are congruent with DeFi-native retail usage, enabled by the lower transaction costs and faster finality on L2s. 
The structural constraint, though conducive to regulatory compliance, limits composability with standard ERC-20 interfaces and precludes permissionless usage; further bifurcating institutional administrative flows from retail DeFi interactions across different layers of the chain stack.

\section{Address Role Predictive Modeling}
We focus on address role predictive modeling within the transactions, inferring (classifying) the financial roles of addresses, i.e., not the identity per se, but the financial or operational role.
Financial role inference helps potentially contextualize address behaviour, differentiate structural actors in token flows, and support future analysis of token activity.
Such financial roles can, in part, be derived from existing address labels.
However, on-chain addresses are overwhelmingly unlabeled. 
Inferring transaction intent, such as distinguishing flows from secondary-market trades, requires role inference from activity traces, which is nontrivial. 
Community-curated sources \cite{duneanalytics2025, arkham-intel, ethplorer2025} and heuristics-based tagging in prior work \cite{makarov2021blockchain, victor2020address, linoy2021anonymizing} provide ex post labels.
These labeling processes are evolving only as addresses gain visibility and social salience, and heuristic methods similarly rely on manually observed patterns, limiting their applicability to new transaction adoption.
Graph-based machine learning models offer an improved approach by enabling automatic, ex ante role inference from activity traces, approximating financial roles directly from transactions that can be updated continuously \cite{zhou2023cryptocurrency, khan2022graph}.
Such automation supplies behaviourally grounded priors that can guide interpretation, narrow the search space for manual investigation, to reduce the cost required for future comprehensive labeling.
Concretely, we advocate for the inference of financial behaviour-grounded roles: financial controllers of treasury, execution (arbitrage) bots, and retail traders, as proxy labels. 
These classes map to distinct economic functions: treasury issuance or redemption liquidity, arbitrageurs, and retail participation, enabling cleaner flow separation and measurement of adoption. 
The capacity for such analysis has been widely needed across academic, industry, and policy communities (e.g., risk, compliance workflows, market monitoring) \cite{makarov2020trading, carapella2023tokenization, makarov2020trading}, motivating our focus on these three classes of roles.
Treasuries refer to multisignature-controlled addresses or custodial vaults (e.g., Gnosis Safe contracts, DAO-managed wallets) that act as long-term capital reserves, execute protocol expenditures, or manage liquidity across chains.
Execution bots engage in high-frequency, low-latency strategies (e.g., sandwiching, MEV extraction, flash-loan arbitrage, Flashbots relaying) whose signatures deviate from standard participation. Aggressive MEV bots can distort price discovery and congest networks, motivating explicit identification, as shown in prior studies \cite{niedermayer2024detecting, lehar2023battle}.
In contrast, retail traders include externally owned accounts (EOAs) that interact primarily with DEXs, NFT marketplaces, or aggregator routers. 

\subsection{U.S. Treasury Tokens Dataset Annotation}
\label{sec:samples_and_labels}
We aggregate transactions from three representative tokenized real-world assets (RWAs): USDY (Ethereum), BENJI (excluding Stellar), and BUIDL.  
These RWA U.S. Treasury tokens were chosen for their active cross-chain circulation, institutional provenance \cite{madhavji2025real}, and relatively more dense community-labeled annotations, making them well-suited for semi-supervised role inference.
We extract the naming tags for each address using Dune SQL from the metadata of Dune \textit{labels.addresses}  \cite{dune_labels_addresses}: a curated repository of community-submitted and platform-extracted multi-chain address labels. 
Each entry includes fields such as \textit{name}, \textit{address}, \textit{blockchain}, \textit{source}, \textit{contributor}, etc.
We utilize the \textit{name} field in the table, which contains the naming tags, then apply a set of regular-expression-based rules over the name field to assign coarse-grained functional roles. 
These rules approximate the aforementioned financial behavioural classes. 
Table~\ref{tab:labeling-rules} summarizes the pattern-matching logic used to infer each role: 
In total, the sampled dataset contains 10,055 transactions between 2023-09-18 and 2025-04-23, covering 815 unique addresses observed as senders or recipients. 
Using this labeling scheme, we label the address set into four classes: 520 \textit{Trader}, 33 \textit{Bot}, 44 \textit{Treasury}, and 218 \textit{Other}. These labels serve as training targets for our predictive modeling pipeline, also allowing us to evaluate role inference performance under supervision.

\begin{table}[t!]
\centering
\caption{Regular expression used for coarse address role labeling.}
\resizebox{0.49\textwidth}{!}{%
\begin{tabular}{ll}
\textbf{Class} & \textbf{Regex } \\
\toprule
\textbf{\textit{Trader}} & \texttt{dex trader}, \texttt{aggregator trader}, \texttt{nft trader}, \texttt{daily trader}, \texttt{number of DEXs traded} \\
\textbf{\textit{Bot}} & \texttt{Sandwich Attacker}, \texttt{Arbitrage}, \texttt{MEV}, \texttt{Flashloan}, \texttt{Flashbots} \\
\textbf{\textit{Treasury}} & \texttt{Safe}, \texttt{Gnosis Safe}, \texttt{Multisig}, \texttt{DAO Treasury}, \texttt{Vault}, \texttt{Zerion Multisig} \\
\textbf{\textit{Other}} & \textit{No match with above patterns} \\
\bottomrule
\end{tabular}}
\label{tab:labeling-rules}
\end{table}

\subsection{Address Representation Learning and Role Inference}
\begin{figure}[t!]
  \centering
  \includegraphics[width=0.99\linewidth]{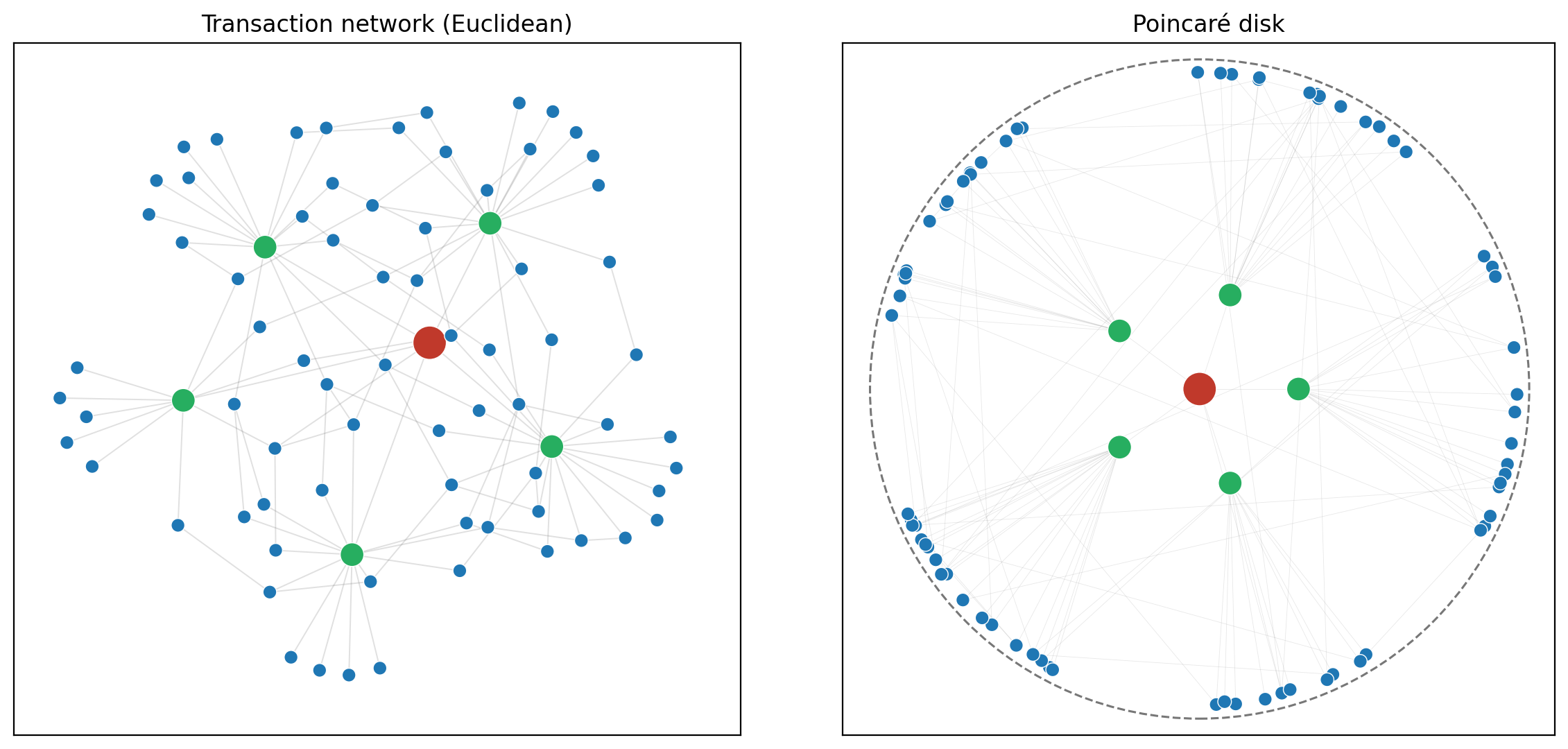}
  \caption{%
  Illustrative example of a blockchain transaction graph and its corresponding Poincaré disk embedding.  
  Nodes emulate three‐tier addresses: the red address anchors global liquidity, the green addresses relay funds and manage liquidity pools, and a ring of peripheral trader blue addresses engage sporadically with the core.  
  Embedding the same adjacency structure in the negatively curved Poincaré model separates tiers by geodesic radius.
  }
  \label{fig:hyperbolic-vs-euclidean}
\end{figure}

U.S. Treasury token transactions exhibit latent financial hierarchy: central actor nodes (e.g., treasuries, issuers, custodians) initiate and coordinate flows, while peripheral nodes (e.g., traders, bots) interact with the core.
The negative curvature of hyperbolic space provides a proper inductive bias for embedding such structure (demonstrated in prior work \cite{liu2019hyperbolic}), enabling representations that naturally separate high-degree core nodes near the origin from low-degree boundary nodes along exponentially expanding geodesics.
Hence, nodes residing at small hyperbolic radii embody \emph{high hierarchy}, systemically close to the coordinate origin, possibly multisig treasuries, protocol routers, or liquidity hubs, whereas nodes with larger radii reflect lower-hierarchy participants in the transaction graph.
A formal derivation showing that hyperbolic radius encodes hierarchical depth in tree-like graphs is given in Section~\ref{sec:hyperbolic-radius}, with illustrative support from Figure~\ref{fig:hyperbolic-vs-euclidean} and an empirical example in Figure~\ref{fig:role-radius}.
Therefore, we propose a hyperbolic node-level representation learning method combined with a feedforward neural network that integrates transactional features, metadata-driven features including Liquidity-to-Average Ratio (LAR), and hyperbolic (Poincaré) geometry embeddings. 
The pipeline first encodes latent hierarchy via Poincaré distance-based optimization, then augments each node with hierarchical depth statistics and optional topological features, processed by a neural network classifier to infer address roles.

\subsubsection{Poincaré Node Representation Learning}
\label{sec:node_representation_learning}
Each data point in our framework corresponds to a node $v \in V$ within a token transaction graph $G = (V, E)$, where nodes represent blockchain addresses and directed edges $(u, v) \in E$ denote value transfers. 
The objective is to learn a high-dimensional representation (embedding vector) $\mathbf{z}_v \in \mathbb{B}^d$ for each address $v$, where $\mathbb{B}^d$ is the $d$-dimensional Poincaré ball: a Riemannian manifold of constant negative curvature:

$$
\mathbb{B}^d = \left\{ \mathbf{z} \in \mathbb{R}^d \;:\; \|\mathbf{z}\| < 1 \right\}.
$$
The manifold is equipped with a Riemannian metric that scales the Euclidean inner product via a position-dependent conformal factor:
$$
g_{\mathbf{z}} = \lambda_{\mathbf{z}}^2 g^{\mathrm{E}}, \qquad \lambda_{\mathbf{z}} = \frac{2}{1 - \|\mathbf{z}\|^2},
$$
where $g^{\mathrm{E}}$ denotes the standard Euclidean metric tensor and $\lambda_{\mathbf{z}}$ diverges as $\|\mathbf{z}\| \to 1$, i.e., as points approach the boundary of the ball. 
The induced geodesic distance between any two points $\mathbf{z}_u, \mathbf{z}_v \in \mathbb{B}^d$ is defined by:
$$
d_{\mathbb{B}}(\mathbf{z}_u, \mathbf{z}_v) = \operatorname{arcosh} \left( 1 + \frac{2 \|\mathbf{z}_u - \mathbf{z}_v\|^2}{(1 - \|\mathbf{z}_u\|^2)(1 - \|\mathbf{z}_v\|^2)} \right),
\label{eq:poincare-distance}
$$
where $\|\cdot\|$ is the Euclidean norm. This metric strongly separates nodes at different depths of the hierarchy, with distances growing rapidly near the boundary.
Möbius addition is calculated when performing updates while preserving manifold structure, a closed-form generalization of vector translation compatible with hyperbolic geometry:
$$
\mathbf{z}_u \oplus \mathbf{z}_v =
\frac{(1 + 2 \langle \mathbf{z}_u, \mathbf{z}_v \rangle + \|\mathbf{z}_v\|^2)\mathbf{z}_u + (1 - \|\mathbf{z}_u\|^2)\mathbf{z}_v}
{1 + 2 \langle \mathbf{z}_u, \mathbf{z}_v \rangle + \|\mathbf{z}_u\|^2 \|\mathbf{z}_v\|^2}.
$$
Each update step is followed by a projection back onto the open ball to ensure validity:
$$
\operatorname{proj}(\mathbf{z}) = \mathbf{z} \cdot \min\left(1, \frac{1 - \varepsilon}{\|\mathbf{z}\| + \varepsilon} \right),
$$
with $\varepsilon > 0$ for numerical stability near the boundary.

We train embeddings via Riemannian stochastic gradient descent (RSGD) over a contrastive objective. Given a positive edge $(i, j^+) \in E$ and a negative sample $j^- \sim \operatorname{Unif}(V)$, we minimize the hinge-based loss:
$$
\mathcal{L}_{\mathrm{c}} = \frac{1}{|E|} \sum_{(i, j^+) \in E}
\left[ d_{\mathbb{B}}(\mathbf{z}_i, \mathbf{z}_{j^+}) - d_{\mathbb{B}}(\mathbf{z}_i, \mathbf{z}_{j^-}) + \gamma \right]_+,
\label{eq:contrastive-loss}
$$
where $\gamma > 0$ is a margin hyperparameter and $[\cdot]_+ = \max(0, \cdot)$ denotes the hinge operator.
To align learned radii with graph-theoretic centrality, we regularize embedding norms against normalized degree:
$$
\mathcal{L}_{\mathrm{r}} = \frac{1}{|V|} \sum_{v \in V}
\left( \|\mathbf{z}_v\| - \left(1 - \frac{\deg(v)}{\max_{u \in V} \deg(u)} \right) \right)^2,
\label{eq:radial-regularization}
$$
encouraging high-degree nodes to concentrate near the center and peripheral actors to occupy the hyperbolic fringe.
The total training objective combines contrastive and curvature-alignment terms:
$$
\mathcal{L} = \mathcal{L}_{\mathrm{c}} + \beta \cdot \mathcal{L}_{\mathrm{r}},
\label{eq:total-loss}
$$
where $\beta$ (default 0.1) controls the strength of radial regularization. 
The objective is minimized over epochs using intrinsic gradients computed in the Riemannian manifold, followed by Möbius updates and projection.
After optimization, each node $v$ is assigned a $\mathbf{z}_v \in \mathbb{B}^{64}$, a 64-dimensional hyperbolic embedding that encodes its topological position and connectivity within the latent hierarchy of the transaction graph.

\vspace{0.1em}
\noindent \textbf{Refinement via Liquidity-to-Average Ratio (LAR)}.
\label{sec:lar}
While Poincaré embeddings encode topological position in the latent transactional hierarchy, other features are related to token flow transferred within transactions, i.e., the temporal liquidity behaviour that distinguishes roles such as passive holders, programmatic bots, or asymmetric treasuries. 
To incorporate such information, we define the following over nodes computed on transaction flows.
\textit{Definition: Liquidity-to-Average Ratio (LAR)}:
\label{def:lar}
Let $(u, v) \in E$ denote a directed transaction edge, and let $\mu_{uv}$ and $\sigma_{uv}$ denote the mean and standard deviation of transfer values from $u$ to $v$ within a time window $[t, t + \Delta]$. Let $\texttt{in}(v)$ and $\texttt{out}(v)$ denote the total inflow and outflow of $v$ over the same interval. Then the Liquidity-to-Average Ratio (LAR) for $(u, v)$ is defined as:
$$
\mathrm{LAR}_{u \to v} =
\frac{\sigma_{uv}}{\mu_{uv} + \epsilon} \cdot 
\left( 1 + \frac{\sum \texttt{in}(v)}{\sum \texttt{out}(v) + \epsilon} \right),
$$
where $\epsilon > 0$ is a smoothing constant to ensure numerical stability.
$\mathrm{LAR}_{u \to v}$ captures local volatility and directional imbalance: the first term reflects normalized transaction variance; the second penalizes outflow-dominant behaviour. 
Nodes with high incoming volume and irregular flow patterns exhibit elevated LAR values.
To integrate this signal into the hyperbolic embedding space, we aggregate edge-level LAR values into node-level weights. 
Let $\log(\texttt{val}_i)$ denote the log-transformed total value received by node $i$, and $\log(\texttt{LAR}_i)$ the log of its average incident LAR. We compute z-scored forms:
$$
z^{\mathrm{val}}_i = \frac{\log(\texttt{val}_i) - \mu_{\log(\texttt{val})}}{\sigma_{\log(\texttt{val})}}, \quad
z^{\mathrm{lar}}_i = \frac{\log(\texttt{LAR}_i) - \mu_{\log(\texttt{LAR})}}{\sigma_{\log(\texttt{LAR})}},
$$
and define node trust as the sigmoid of their difference: $\tau_i = \sigma(z^{\mathrm{val}}_i - z^{\mathrm{lar}}_i)$.
These trust weights modulate a refinement step in hyperbolic space. For each node $i$, let $\mathcal{N}(i)$ denote its neighbours in $G$. Define the trust-weighted tangent update:
$$
\mathbf{t}_i = \sum_{j \in \mathcal{N}(i)} \alpha_{ij} \cdot \log_0(\mathbf{z}_j), \quad
\alpha_{ij} = \frac{\tau_j}{\sum_{k \in \mathcal{N}(i)} \tau_k},
$$
where $\log_0(\cdot)$ is the logarithmic map at the origin. The new embedding is then updated via Möbius exponential map:
$$
\mathbf{z}_i \leftarrow \operatorname{proj} \left( \tanh\left(\frac{\|\mathbf{t}_i\|}{2}\right) \cdot \frac{\mathbf{t}_i}{\|\mathbf{t}_i\| + \delta} \right),
$$
with $\delta > 0$ ensuring stability near $\|\mathbf{t}_i\| = 0$. We apply this refinement for $R=3$ steps, with early stopping if embeddings converge.

\textit{Observation}.
Nodes with low liquidity and irregular behaviour are downweighted during smoothing, preserving geometric sharpness for high-confidence addresses (e.g., well-funded treasuries), while dampening noise from sparsely observed actors.

\subsubsection{Hyperbolic Radius as a Proxy for Latent Hierarchy}
\label{sec:hyperbolic-radius}

Let $\mathbb{T}_k$ denote a $k$-ary tree, and consider a mapping $\phi: \mathbb{T}_k \rightarrow \mathbb{B}^d$ from a tree into the $d$-dimensional Poincaré ball $\mathbb{B}^d = \{ \mathbf{z} \in \mathbb{R}^d \mid \|\mathbf{z}\| < 1 \}$.

\begin{figure}[t]
    \centering
    \includegraphics[width=0.68\linewidth]{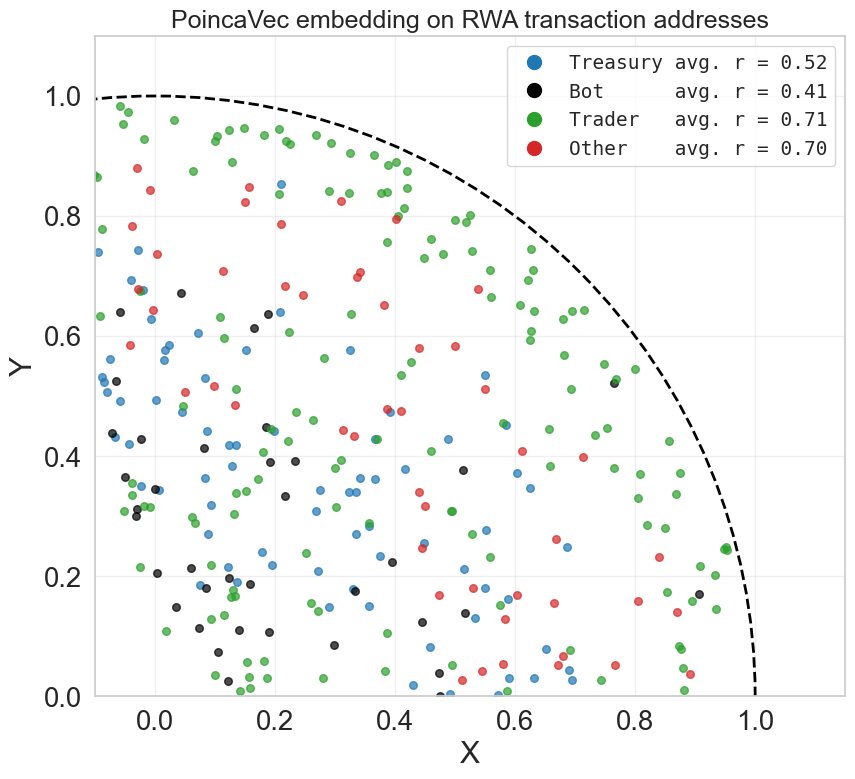}
    \caption{Illustration of PoincaVec embedding of addresses colored by labels from our U.S. Treasury dataset. Smaller hyperbolic radius indicates higher hierarchy (closer to the origin). Roles like \textsc{Treasury} and \textsc{Bot} appear near the origin, while \textsc{Trader} and \textsc{Other} are peripheral, near the edge.}
    \label{fig:role-radius}
\end{figure}

Let $u$ be a node at depth $h$ in the tree (i.e., $h$ hops from the root). Suppose $\phi$ maps nodes along a fixed geodesic direction such that the hyperbolic distance from the origin encodes depth:
$$
r_h := \|\phi(u)\| = \tanh\left( \frac{h \cdot \ell}{2} \right),
$$
where $\ell > 0$ is the fixed geodesic step length in the hyperbolic metric. 
The detailed derivation and proof of the corresponding lemma are provided in Appendix \ref{appendix:hyperbolic-depth}.
To empirically illustrate that Poincaré embeddings encode latent hierarchy, we visualize address embeddings in the Poincaré disk, with nodes colored by annotated roles and average hyperbolic radius used as a proxy for structural depth. 
Figure~\ref{fig:role-radius} shows three hundred randomly sampled addresses from the RWA Treasury dataset, visualized in two-dimensional hyperbolic space, demonstrating how hyperbolic radius captures latent hierarchical structure. 

\subsubsection{Hierarchical Radius-based Features}
\label{sec:radius_feature}
To capture the local structure around node $v$, we define its $k$-hop neighbourhood $\mathcal{N}_k(v) \subseteq V$ (excluding $v$) and collect neighbour radii $\{r_u : u \in \mathcal{N}_k(v)\}$. 
From this, we compute an 11-dimensional hierarchical feature vector detailed in Appendix \ref{appx:radius_feature}.

\subsubsection{Neural Network Inferencer}
\label{sec:inferencer}
After obtaining the Poincaré position embedding and hierarchical radius-based features, we train a supervised neural classifier to infer coarse address roles. 
Each node $v \in V$ is represented by a concatenated input vector:
\begin{itemize}
    \item $\mathbf{z}_v \in \mathbb{R}^d$ (constrained to $\mathbb{B}^d$): Poincaré embedding (cf. Section~\ref{sec:node_representation_learning});
    \item $\mathbf{h}_v \in \mathbb{R}^{11}$: hierarchical features computed from local radius-based statistics (cf. Section~\ref{sec:radius_feature});
    \item $\mathbf{r}_v \in \mathbb{R}^{k}$: vector of topology-aware node features derived from truncated random walks over local neighbourhoods (as in DeepWalk embeddings \cite{perozzi2014deepwalk}).
\end{itemize}
These components are concatenated into a unified feature vector:
$$
\mathbf{x}_v = \left[ \mathbf{z}_v \,\middle\|\, \mathbf{r}_v \,\middle\|\, \mathbf{h}_v \right] \in \mathbb{R}^{d + k + 11},
$$
where $d$ is the Poincaré embedding dimension, $k$ is the raw feature dimension.
The neural architecture consists of a two-layer multilayer perceptron (MLP) with batch normalization, ReLU activations, and dropout regularization. Letting $\mathrm{MLP}: \mathbb{R}^{d + k + 11} \rightarrow \mathbb{R}^C$ denote the classifier, the unnormalized logits for node $v$ are:
$$
\hat{\mathbf{y}}_v = \mathrm{MLP}(\mathbf{x}_v) = \mathbf{W}_2 \cdot \mathrm{ReLU}\left( \mathbf{W}_1 \cdot \mathbf{x}_v + \mathbf{b}_1 \right) + \mathbf{b}_2,
$$
where $C$ is the number of target classes.
Training is supervised using the standard cross-entropy loss over weak role labels $\{y_v\}_{v \in \mathcal{D}_{\mathrm{train}}}$:
$$
\mathcal{L}_{\mathrm{clf}} = - \sum_{v \in \mathcal{D}_{\mathrm{train}}} \log \frac{ \exp(\hat{y}_v^{(y_v)}) }{ \sum_{c=1}^{C} \exp(\hat{y}_v^{(c)}) }.
$$
This loss is minimized when the model assigns high probability to the correct class. 
The ground-truth label $y_v$ selects the corresponding logit $\hat{y}_v^{(y_v)}$ from the model's output $\hat{\mathbf{y}}_v$, and the loss penalizes the model when this logit does not dominate the softmax distribution, i.e., when the predicted probability for class $y_v$ is low.

\subsubsection{Experimental Results}
\label{sec:experiment-result}
We evaluate our role classifier using a stratified train/test split with an 80/20 ratio over the set of labeled addresses with the four coarse-grained role classes (cf. Section \ref{sec:samples_and_labels}). 
The model is optimized using the AdamW \cite{loshchilovdecoupled} optimizer with a learning rate of $10^{-3}$ and a weight decay of $0.2$. 
Training is conducted for up to 2k epochs with early stopping based on macro-F\textsubscript{1} score on the validation set, using a patience threshold of 10 epochs. 
The best-performing model checkpoint (by macro-F\textsubscript{1}) is selected for final evaluation. 
We conduct an ablation study using the input features of Poincaré embeddings $\mathbf{z}_v \in \mathbb{R}^{64}$, hierarchical descriptors $\mathbf{h}_v \in \mathbb{R}^{11}$ ($k=1$), and topology-aware node embeddings $\mathbf{r}_v \in \mathbb{R}^{64}$, with feature subsets in combination under various ablation settings.
We compare our model \textit{PoincaVec} against three established node representation baselines: Node2Vec \cite{grover2016node2vec}, Role2Vec \cite{ahmed2019role2vec}, and FeatherNode \cite{rozemberczki2020characteristic}, included for their demonstrated performance on blockchain transaction graphs in prior evaluations~\cite{gangemi2024overview, beres2021blockchain}.
We adopt the same hyperparameter settings as in prior blockchain address representation learning works \cite{poursafaei2021sigtran, beres2021blockchain}, setting the context size $10$, embedding dimension $64$, walk length $5$, number of walks per node $10$ for both Node2Vec and Role2Vec, while FeatherNode is configured Singular Value Decomposition (SVD) iterations 20.
The baseline Node2Vec~\cite{grover2016node2vec} captures basic neighbour homophily-based proximity via biased random walks, while Role2Vec~\cite{ahmed2019role2vec} models structural equivalence by clustering nodes based on topological statistics into roles (types) and learning role-level embeddings where the nodes share the same role obtain the same embedding vector. 
FeatherNode \cite{rozemberczki2020characteristic} incorporates spectral and node-level distributional features from attributed graphs by applying Fourier transforms of empirical distributions to neighbourhood feature aggregations, subsequently compressed using truncated SVD, yielding node embeddings invariant to permutation and sensitive to local graph structure.
Table~\ref{tab:role-results} presents the classification performance of PoincaVec and several baseline embedding methods on the RWA U.S. Treasury dataset. 
PoincaVec (w/ H, w/o T) outperforms all baselines with an F\textsubscript{1} of 0.693. 
Integrating hierarchical radius features with topology-aware node vectors derived from truncated random walks (w/ H, w/ T) yields an F1 score of 0.726. 
These results demonstrate that the PoincaVec embeddings, produced by the proposed method enriched with structural features, provide competitive representational capacity for role inference in blockchain transaction graphs.

\begin{table}[t!]
\centering
\caption{Role classification performance on the RWA U.S. Treasury dataset. H: hierarchical radius-based features; T: topology-aware node features.}
\renewcommand{\arraystretch}{0.85}
\resizebox{0.48\textwidth}{!}{%
\begin{tabular}{lcccc}
\toprule
\textbf{Model} & \textbf{Precision} & \textbf{Recall} & \textbf{F1} & \textbf{Accuracy} \\
\midrule
Node2Vec                          & 0.668 & 0.687 & 0.654 & 0.687 \\
Role2Vec                          & 0.659 & 0.681 & 0.662 & 0.681 \\
FeatherNode                       & 0.407 & 0.638 & 0.497 & 0.638 \\
PoincaVec (w/ H, w/o T)           & 0.690 &  0.712  & 0.693  & 0.712 \\
PoincaVec (w/o H, w/ T)           & 0.735 & 0.736  & 0.716 &  0.736  \\
PoincaVec (w/ H, w/ T)            & 0.757 & 0.748 & \textbf{0.726} & \textbf{0.748} \\
\bottomrule
\end{tabular}
}
\label{tab:role-results}
\end{table}

\subsubsection{Extended Evaluation on Public Blockchain Transaction Datasets}
\label{sec:external-datasets}
To further evaluate the generalization capacity of our PoincaVec architecture beyond the RWA role classification task, we test it on multiple publicly available blockchain transaction graph datasets with labels. Detailed statistics and descriptions of the datasets are in Appendix~\ref{app:datasets}.
These datasets include addresses labeled as fraudulent or scam-related, whose topological irregularities render them particularly amenable to detection using curvature-aware embeddings and hierarchical structural descriptors.
Given that fraud or anomalous addresses often lie structurally at the periphery or occupy non-homophilic positions in the graph, the curvature-aware inductive bias of PoincaVec may help reveal subtle hierarchical distinctions absent in flat Euclidean spaces.

\begin{table}[t!]
\centering
\caption{Role classification performance across four blockchain transaction datasets. }
\label{tab:all-dataset-results}
\renewcommand{\arraystretch}{0.85}
\resizebox{0.48\textwidth}{!}{%
\begin{tabular}{lcccc}
\toprule
\multicolumn{5}{l}{\textsc{Ethereum}} \\
\midrule
\textbf{Model} & \textbf{Precision} & \textbf{Recall} & \textbf{F1} & \textbf{Accuracy} \\
Node2Vec                       & 0.922 & 0.918 & 0.918 & 0.918 \\
Role2Vec                       & 0.923 & 0.923 & 0.923 & 0.923 \\
FeatherNode                    & 0.922 & 0.916 & 0.916 & 0.916 \\
PoincaVec (w/ H, w/o T)        & 0.901 & 0.897 & 0.897 & 0.896 \\
PoincaVec (w/o H, w/ T)        & 0.936 & 0.936 & 0.936 & 0.936 \\
PoincaVec (w/ H, w/ T) & 0.941 & 0.940 & \textbf{0.940} & \textbf{0.940} \\
\midrule
\multicolumn{5}{l}{\textsc{AscendEXHacker}} \\
\midrule
Node2Vec                       & 0.889 & 0.719 & 0.780 & 0.991 \\
Role2Vec                       & 0.829 & 0.675 & 0.728 & 0.989 \\
FeatherNode                    & 0.694 & 0.558 & 0.587 & 0.986 \\
PoincaVec (w/ H, w/o T)        & 0.935 & 0.779 & \textbf{0.840} & \textbf{0.993} \\
PoincaVec (w/o H, w/ T)        & 0.925 & 0.739 & 0.806 & 0.992 \\
PoincaVec (w/ H, w/ T) & 0.897 & 0.735 & 0.794 & 0.992 \\
\midrule
\multicolumn{5}{l}{\textsc{PlusTokenPonzi}} \\
\midrule
Node2Vec                       & 0.996 & 0.990 & 0.993 & 0.996 \\
Role2Vec                       & 0.993 & 0.996 & 0.994 & 0.996 \\
FeatherNode                    & 0.996 & 0.992 & 0.994 & 0.996 \\
PoincaVec (w/ H, w/o T)        & 0.995 & 0.979 & 0.987 & 0.992 \\
PoincaVec (w/o H, w/ T)        & 0.997 & 0.991 & \textbf{0.994} & \textbf{0.996} \\
PoincaVec (w/ H, w/ T) & 0.996 & 0.986 & 0.991 & 0.994 \\
\midrule
\multicolumn{5}{l}{\textsc{Ethereum Classic (ETC)}} \\
\midrule
Node2Vec                       & 0.910 & 0.898 & 0.899 & 0.899 \\
Role2Vec                       & 0.888 & 0.882 & 0.882 & 0.882 \\
FeatherNode                    & 0.927 & 0.926 & 0.926 & 0.926 \\
PoincaVec (w/ H, w/o T)        & 0.910 & 0.898 & 0.899 & 0.900 \\
PoincaVec (w/o H, w/ T)        & 0.921 & 0.913 & 0.914 & 0.914 \\
PoincaVec (w/ H, w/ T) & 0.949 & 0.943 & \textbf{0.945} & \textbf{0.946} \\
\bottomrule
\end{tabular}
}
\end{table}

Table~\ref{tab:all-dataset-results} presents a comparison of benchmark methods and our PoincaVec variants across four blockchain transaction graph datasets with varying structural properties and labels. 
Across all datasets, the full PoincaVec pipeline incorporating both hierarchical radius-based features (H) and topology-aware walk embeddings (T) achieves the highest or near-highest F1 scores, demonstrating generalization to anomalous address classification tasks. 
For the large, sparsely labeled network of Ethereum dataset, PoincaVec (w/ H, w/ T) achieves an F1 of 0.940, outperforming proximity-based baselines such as Node2Vec and structurally driven Role2Vec. 
The \textsc{AscendEXHacker} dataset exhibits an imbalanced role distribution, where only a small fraction of addresses are tagged as direct heist participants, while the majority are exchange-facing services such as DEX routers (e.g., Uniswap). 
In this setting, hierarchical radius-based features (H) offer an inductive prior: heist addresses tend to lie peripherally with sparse connectivity, while services cluster near radius, enabling PoincaVec (w/ H, w/o T) to achieve the highest F1 of 0.840, whereas adding walk-based features reduces performance to 0.794.
Notably, PoincaVec obtained an F1 of 0.945 on the Ethereum Classic (ETC) dataset, demonstrating its capacity to disentangle abnormal node roles. 
These results demonstrate the performance of embedding blockchain transaction graphs in hyperbolic space, where latent hierarchies and peripheral node behaviours, which are common in fraud and scam contexts, can be captured more compactly in our proposed method.
Additional evaluation figures and detailed metric analyses are provided in Appendix \ref{appx:extended-eval}, including full Precision@k curves and prevalence-sensitivity visualizations across all datasets.

\section{Conclusion}
This study presents the first transaction-level analysis of tokenized U.S. Treasuries across multiple blockchain networks. 
By decoding contract function usage, we identify dominant operational roles and reveal distinct patterns between institutional and retail participants, and provide a quantitative account of how institutional and retail participants engage with tokenized U.S. Treasuries across chains and transactional contexts.
We proposed a Poincaré embedding-based model that enables role inference and demonstrates improved performance both on our RWA dataset and when generalizing to external anomalous address classification tasks.

\bibliographystyle{ieeetran}
\bibliography{reference.bib}

\appendices

\begin{figure*}[htbp!]
    \centering
    \includegraphics[width=0.333\textwidth]{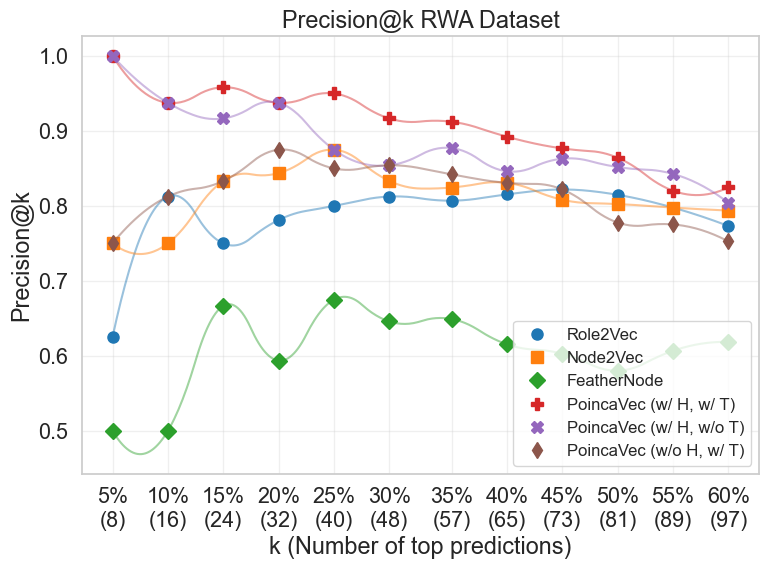} \hspace{-3mm}
    \includegraphics[width=0.333\textwidth]{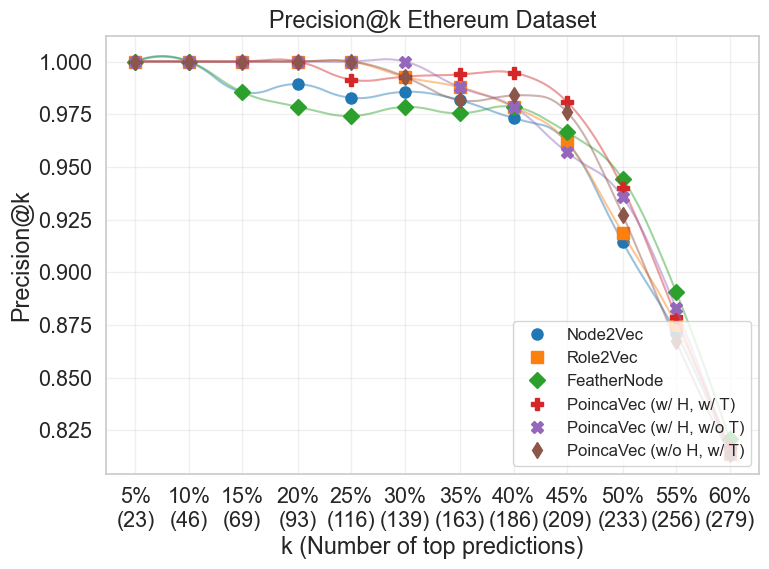} \hspace{-3mm} 
    \includegraphics[width=0.333\textwidth]{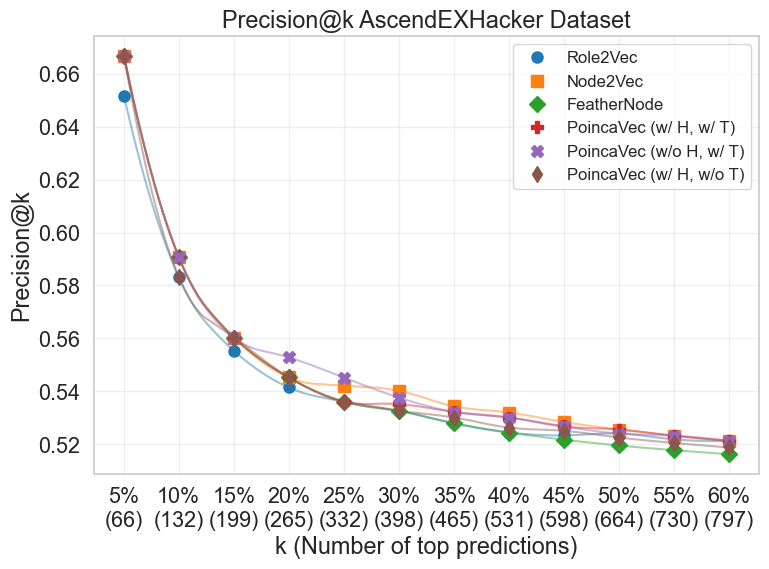} \hspace{-3mm}
    \includegraphics[width=0.333\textwidth]{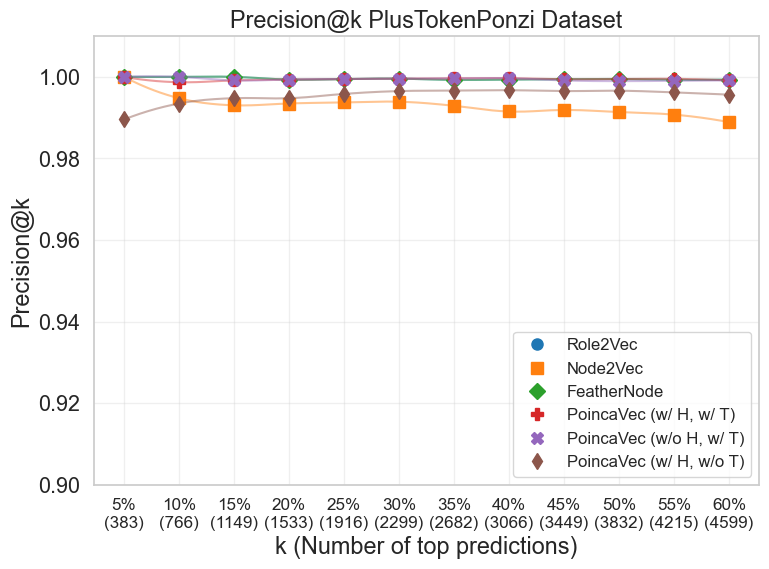} \hspace{-3mm}
    \includegraphics[width=0.333\textwidth]{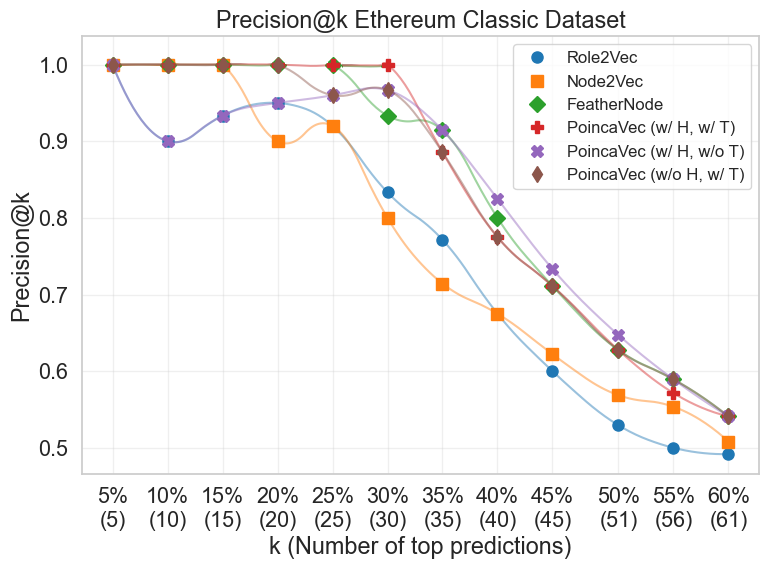} \hspace{-1mm}
    \caption{Precision@k curves on all five datasets, evaluated at k ranges from 0.05 to 0.60 in increments of 0.05 of the test set. Our proposed \textsc{PoincaVec} (w/ H, w/ T) variant demonstrates strong top-$k$ ranking performance across all the datasets, demonstrating the capacity to detect anomalous addresses of various blockchain transaction datasets.}
    \label{fig:precision-k}
\end{figure*}

\section{Datasets details}
\label{app:datasets}
\noindent \textbf{Ethereum Transaction}~\cite{wu2020phishers}
An Ethereum transaction dataset constructed from historical blockchain records, containing 2,973,489 addresses and 13,551,303 edges. A total of 1,165 addresses are labeled as illicit based on curated forensic data, while the rest remain unlabeled. This dataset has an average node degree of 4.56.

\noindent \textbf{AscendEXHacker}~\cite{EthereumHeist}
A network extracted via Etherscan's Heist label, focusing on the 2021 AscendEX exchange exploit. It contains 6,642 addresses across 29,074 transactions. Ground-truth annotations label 84 addresses as direct heist participants, while 638 addresses are identified as DEXs or Uniswap-related service accounts. 

\noindent \textbf{PlusTokenPonzi}~\cite{EthereumHeist}
A transaction dataset linked to the PlusToken Ponzi, one of the largest crypto scams. It comprises 34,521 unique accounts and 58,049 transactions, of which 30,782 addresses are explicitly identified as scam participants. 

\noindent \textbf{Ethereum Classic}~\cite{al2020labeled}
A labeled transaction network from the Ethereum Classic (ETC) network, compiled using EtherscamDB scam reports. The dataset comprises 73,034 nodes and 71,250 edges, with 169 addresses labeled as scammers based on crowdsourced and externally verified annotations.

\section{Hierarchical Radius-based Features Vector}
\label{appx:radius_feature}
To incorporate latent hierarchy from hyperbolic space, we associate each node $v \in V$ with a Poincaré embedding $\mathbf{z}_v \in \mathbb{R}^d$ satisfying $\|\mathbf{z}_v\| < 1$. The node's hierarchical depth is defined via its \emph{hyperbolic radius}:
$$
r_v = 2 \cdot \tanh^{-1}(\|\mathbf{z}_v\|),
$$
which reflects the geodesic distance from the origin and serves as a proxy for depth in the hierarchy, i.e., nodes closer to the origin are higher in the hierarchy, while those near the boundary are deeper.
From the neighbour radii, we compute an 11-dimensional hierarchical feature vector $\mathbf{h}_v \in \mathbb{R}^{11}$ comprising the following components:

\vspace{0.5em}
\begin{tabular}{@{}ll@{}}
$r_v$         & Absolute depth (node's self radius) \\
$\mu$         & Mean of neighbour radii: $\mu = \frac{1}{|\mathcal{N}_k(v)|} \sum\limits_{u \in \mathcal{N}_k(v)} r_u$ \\
$\sigma$ & \makecell[l]{Standard deviation of neighbour radii: \\ $\sigma = \sqrt{\frac{1}{|\mathcal{N}_k(v)|} \sum\limits_{u \in \mathcal{N}_k(v)} (r_u - \mu)^2}$} \\
$\alpha$ & \makecell[l]{Fraction of neighbours deeper than $v$: \\ $\alpha = \frac{1}{|\mathcal{N}_k(v)|} \sum\limits_{u} \mathbb{I}(r_u > r_v)$} \\
$\beta$ & \makecell[l]{Fraction of neighbours shallower than $v$: \\ $\beta = \frac{1}{|\mathcal{N}_k(v)|} \sum\limits_{u} \mathbb{I}(r_u < r_v)$} \\
$\delta$      & Minimum relative depth: $\delta = \min\limits_{u \in \mathcal{N}_k(v)} (r_u - r_v)$ \\
$\Delta$      & Maximum relative depth: $\Delta = \max\limits_{u \in \mathcal{N}_k(v)} (r_u - r_v)$ \\
$\mathbf{b}_v$ & 4-bin histogram of relative radii $(r_u - r_v)$ \\
 & over range $[-1, 1]$
\end{tabular}
\vspace{0.5em}

Altogether, the hierarchical feature vector is given by:
$$
\mathbf{h}_v = [r_v, \mu, \sigma, \alpha, \beta, \delta, \Delta \;|\; \mathbf{b}_v \in \mathbb{R}^{4}] \in \mathbb{R}^{11},
$$
providing a compact representation that summarizes both the absolute radial position of node $v$ and the statistical distribution of depths in its local hyperbolic neighbourhood.

\section{Hyperbolic Depth Derivation}
\label{appendix:hyperbolic-depth}

\textit{Lemma}.
The hyperbolic radius $r_h$ is a strictly increasing function of tree depth $h$.

\begin{proof}
Since $\tanh(x)$ is a strictly increasing function for $x > 0$, and $h \mapsto h\ell/2$ is linear and positive for $h \geq 0$, we have:
$$
\frac{d}{dh} r_h = \frac{d}{dh} \tanh\left( \frac{h \ell}{2} \right) = \frac{\ell}{2} \cdot \text{sech}^2\left( \frac{h \ell}{2} \right) > 0.
$$
Therefore, $r_h$ increases monotonically with $h$.
\end{proof}
\textit{Corollary}.
Under tree-consistent Poincaré embeddings, node radius $\|\mathbf{z}_v\|$ can be interpreted as a continuous proxy for latent hierarchical depth.

\section{Extended Evaluation Metrics}
\label{appx:extended-eval}
\subsubsection{Evaluation of Ranking Performance via Precision@k}
\label{sec:precision-k}

To assess ranking fidelity, we evaluate Precision@k over a range of thresholds $k \in \{\,0.05 \cdot n \mid n = 1, ...,  12\,\} \cdot |\text{test set}|$, defined as fractions of the test set size.
We evaluate top-$k$ ranking performance using Precision@k on all five datasets. 
As shown in Figure~\ref{fig:precision-k}, our proposed \textsc{PoincaVec} variants consistently rank positive instances more effectively than baselines across a range of thresholds. 
In datasets such as \textsc{RWA} and \textsc{AscendEXHacker}, \textsc{PoincaVec (w/ H, w/ T)} delivers substantial gains in early precision, confirming its suitability for skewed detection settings. 
On denser class datasets such as \textsc{PlusTokenPonzi}, where nearly all methods saturate near-perfect precision, the performance of the proposed method remains stable and competitive.
These results suggest that PoincaVec inductively improves F1-oriented retrieval without compromising ranking precision.

\subsubsection{Prevalence-Sensitivity of Performance Improvements}
\label{sec:prevalence-sensitivity}
\begin{figure}[htbp!]
    \centering
    \includegraphics[width=\linewidth]{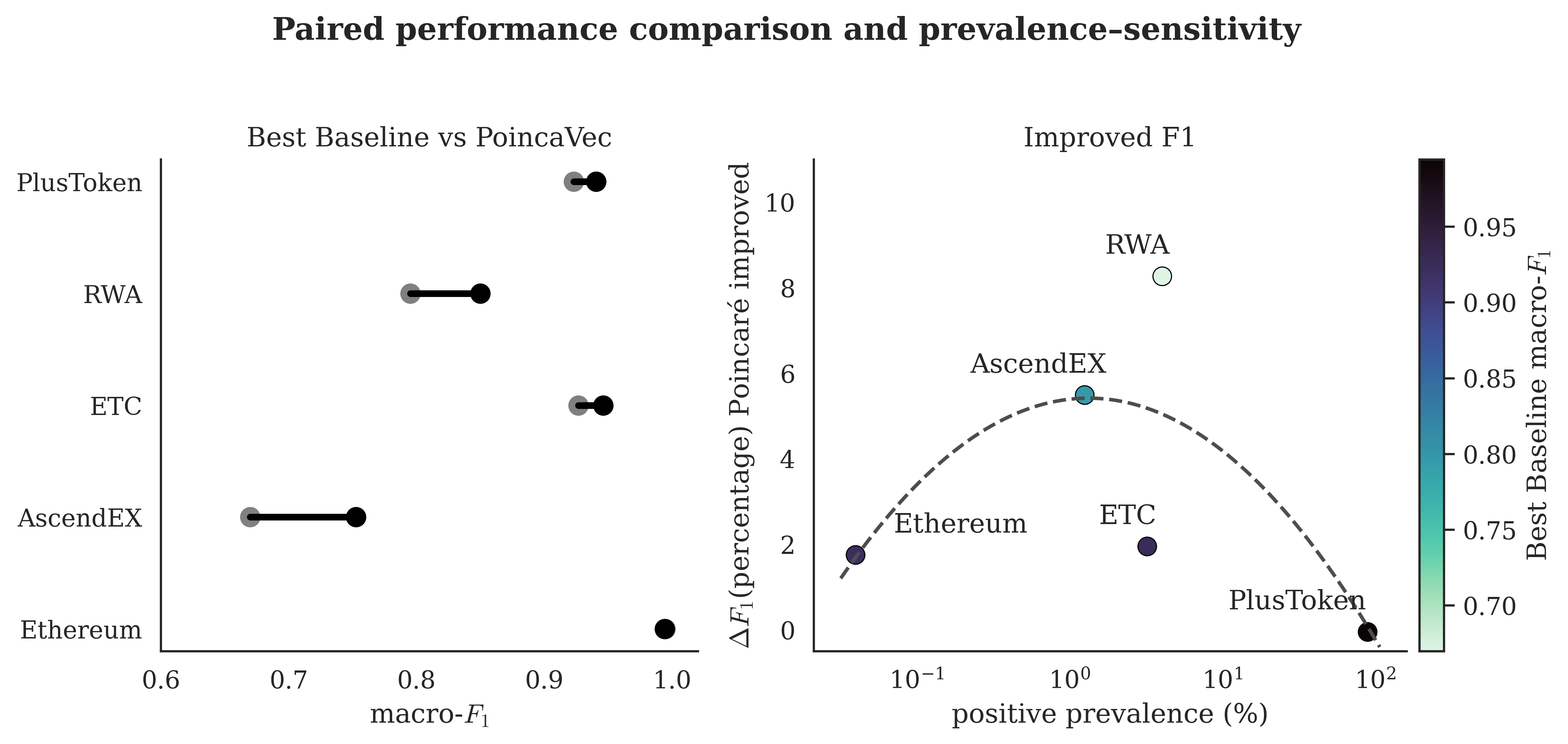}
    \caption{\textbf{Paired performance comparison and prevalence-sensitivity.}
    \textbf{Left:} Macro F1 scores for the best baseline vs. PoincaVec across all five datasets.
    \textbf{Right:} Improvement in macro F1 (percent) plotted against positive prevalence. 
    Color encodes best baseline performance to contextualize potential uplift. 
    The dashed line shows a concave fit, with relatively more improvement at intermediate prevalence levels.}
    \label{fig:prevalence-sensitivity}
\end{figure}

To evaluate the effect of positive class prevalence, we assess prevalence-sensitivity by relating the performance improvement of each dataset to the conducted experiment.
Figure~\ref{fig:prevalence-sensitivity} summarizes two aspects: the paired macro-$F_1$ gains of our method over the best baseline (left), and how those gains vary with positive prevalence (right). 
The left panel visualizes absolute improvements for each dataset, confirming that our proposed model consistently outperforms the baselines on both our collected RWA U.S. Treasury transaction dataset and all other public blockchain transaction datasets.
The right panel plots $\Delta F_1$ uplift against prevalence (log scale), revealing a concave pattern: improvements peak at intermediate prevalence and diminish at both extremes. 
These results indicate that our method achieves the greatest improvements when the positive class is neither too rare nor too dominant, i.e., under moderate imbalance, particularly when the baseline model struggles to separate roles in scenarios where class boundaries are less distinct and prevalence alone does not strongly determine separability.

\textbf{Ethical Considerations}
This study uses only publicly available, pseudonymous blockchain transaction data and performs no linkage to real-world identities; we neither collect nor expose personally identifiable information. All analyses are conducted on‐chain transaction graphs; results are reported in aggregate and we release no per-address classifications, limiting targeting risks. The role-inference model is probabilistic tool for on-chain addresses analyzing and should not be treated as definitive evidence of illicit behaviour. We will publish code that reproduces aggregate findings. Given these safeguards, we think that the transparency benefits for market participants and researchers in the blockchain, and web for good community.

\end{document}